\documentclass[10pt,twocolumn]{article}
\usepackage{times}
 \pdfoutput=1 
\usepackage{graphicx}
\graphicspath{{./figures/}}
\DeclareGraphicsExtensions{.pdf,.jpeg,.png}
\usepackage{balance}  

\PassOptionsToPackage{hyphens}{url}
\usepackage[hyphenbreaks]{breakurl}

\usepackage[capitalise]{cleveref}

\usepackage{listings}
\usepackage{color}
\usepackage{csquotes}
\usepackage{makecell}

\definecolor{dkgreen}{rgb}{0,0.6,0}
\definecolor{gray}{rgb}{0.5,0.5,0.5}
\definecolor{mauve}{rgb}{0.58,0,0.82}

\begin{document}


\title{Stocator:  A High Performance Object Store Connector for Spark}

\author{Gil Vernik$^1$, Michael Factor$^1$, Elliot K. Kolodner$^1$\\ Pietro Michiardi$^2$, Effi Ofer$^1$ and Francesco Pace$^2$\\
	\small {\em  $^1$IBM Research -- Haifa \quad $^2$Eurecom} \\ 
	\small {\em $^1$\{gilv,factor,kolodner,effio\}@il.ibm.com \quad $^2$ \{michiardi,pace\}@eurecom.fr}
	\\ [2mm]
	\small Submission Type: Research
}

\date{}

\maketitle

\begin{abstract}
We present Stocator, a high performance object store connector for
Apache Spark, that takes advantage of object store semantics. Previous
connectors have assumed file system semantics, in particular,
achieving fault tolerance and allowing speculative execution by
creating temporary files to avoid interference between worker threads
executing the same task and then renaming these files. Rename is not a
native object store operation; not only is it not atomic, but it is
implemented using a costly copy operation and a delete. Instead our
connector leverages the inherent atomicity of object creation, and by
avoiding the rename paradigm it greatly decreases the number of
operations on the object store as well as enabling a much simpler
approach to dealing with the eventually consistent semantics typical
of object stores. We have implemented Stocator and shared it in open
source. Performance testing shows that it is as much as 18 times
faster for write intensive workloads and performs as much as 30 times
fewer operations on the object store than the legacy Hadoop
connectors, reducing costs both for the client and the object storage
service provider.
\end{abstract}

\section{Introduction}
Data is the natural resource of the 21st century. It is being produced
at dizzying rates, e.g., for genomics by sequencers, for healthcare
through a variety of imaging modalities, and for Internet of Things
(IoT) by multitudes of sensors. 
This data increasingly resides in cloud object stores,
such as AWS S3\cite{s3}, Azure Blob storage\cite{azure}, and IBM Cloud
Object Storage\cite{ibm-cloud-object-storage},
which are highly scalable distributed cloud storage systems that offer
high capacity, cost effective storage.
But it is not enough just to store data;
we also need to derive value from it,
in particular, through analytics engines such as Apache Hadoop\cite{hadoop} and
Apache Spark\cite{spark}.
However, these highly distributed analytics engines were originally designed
work on data stored in HDFS (Hadoop Distributed File System) where the
storage and processing are co-located in the same server cluster.
Moving data from object storage to HDFS in
order to process it and then moving the results back to object storage
for long term storage is inefficient.
In this paper we present Stocator\cite{stocator-blog}, a high
performance storage connector, that enables Hadoop-based analytics
engines to work directly on data stored in object storage systems.
Here we focus on Spark; our work can be extended to work with the
other parts of the Hadoop ecosystem.

Current connectors to object stores for Spark, e.g., S3a\cite{aws-s3} and the Hadoop Swift Connector\cite{hadoop-swift} are notorious for their poor performance\cite{pace_2016} for write workloads and sometimes leaving behind temporary objects that do not get deleted.
The poor performance of these connectors follows from their assumption of file system semantics, a natural assumption given that their model of operation is based on the way that Hadoop interacts with its original storage system, HDFS\cite{hdfs}.
In particular, Spark and Hadoop achieve fault tolerance and enable speculative execution by creating temporary files and then renaming these files. This paradigm avoids interference between threads doing the same work and thus writing output with the same name. Notice, however, that rename is not a native object store operation; not only is it not atomic, but it must be implemented using a costly copy operation, followed by a delete.

Current connectors can also lead to failures and incorrect executions because the list operation on containers/buckets is eventually consistent. EMRFS\cite{emrfs-blog} from Amazon and S3mper\cite{s3mper-blog} from Netflix overcome eventual consistency by storing file metadata in DynamoDB\cite{aws-dynamodb}, an additional strongly consistent storage system separate from the object store. A similar feature called S3Guard\cite{s3guard-slides} that also requires an additional strongly consistent storage system is being developed by the Hadoop open source community for the S3a connector. Solutions like these, which require multiple storage systems, are complex and can introduce issues of consistency between the stores. They also add cost since users must pay for the additional strongly consistent storage.

Others have tried to improve the performance of object store connectors, e.g., the DirectOutputCommitter\cite{direct-output-committer} for S3a introduced by Databricks, but have failed to preserve the fault tolerance and speculation properties of the temporary file/rename paradigm. There are also recommendations in the Hadoop open source
community to abandon speculation and employ an
optimization\cite{file-output-committer-2} that renames files to their
final names when 
tasks complete (commit) instead of waiting for the completion of the
entire job. However, incorrect executions, though rare, can still
occur even with speculation turned off due to the eventually
consistent list operations employed at task commit to determine which
objects to rename. 

In this paper we present a high performance object store connector for
Apache Spark that takes full advantage of object store semantics,
enables speculative execution and also deals correctly 
with eventual consistency.
Our connector
eliminates the rename paradigm by writing each output object to its
final name.
The name includes both the part number and the attempt
number, so that multiple attempts to write the same part due to
speculation or fault tolerance use different object names.
Avoiding rename also removes the necessity to execute list operations
to determine which objects to rename at task and job commit,
so that a Spark job writes all of the parts constituting its output
dataset correctly despite eventual consistency.
This reduces the issue of eventual consistency to ensuring that a subsequent
job correctly determines the constituent parts when it reads the
output of previous jobs.
Accordingly
we extend an already existing success indicator object written at the
end of a Spark job to include a manifest to indicate the part names
that actually compose the final output. A subsequent job reads the
indicator object to determine which objects are part of the
dataset. 
Overall, our approach increases performance by greatly decreasing the number of
operations on the object store 
and ensures correctness despite eventual consistency by greatly
decreasing complexity. 

Our connector also takes advantage of HTTP Chunked Transfer Encoding to stream the data being written to the object store as it is produced, thereby avoiding the need to write objects to local storage prior to being written to the object store.

We have implemented our connector for the OpenStack Swift API\cite{openstack-swift-api} and shared it in open source\cite{stocator-git}. We have compared its performance with the S3a and Hadoop Swift connectors over a range of workloads and found that it executes far less operations on the object store, in some cases as little as one thirtieth of the operations. Since the price for an object store service typically includes charges based on the number of operations executed, this reduction in the number of operations lowers the costs for clients in addition to reducing the load
on client software. It also reduces costs and load for the object store provider since it can serve more clients with the same amount of processing power. Stocator also substantially increases performance for Spark workloads running over object storage, especially for write intensive workloads, where it is as much as 18 times faster.

In summary our contributions include:
\begin{itemize}
	\item The design of a novel storage connector for Spark that leverages object storage semantics, avoiding costly copy operations and providing correct execution in the face of faults and speculation.
	\item A solution that works correctly despite the eventually consistent semantics of object storage, yet without requiring
	  additional strongly consistent storage.
	\item An implementation that has been contributed to open source.
\end{itemize}
Stocator is in production in IBM Analytics for Apache Spark, a Bluemix
service, and has enabled the SETI project to perform computationally intensive
Spark workloads on multi-terabyte binary signal files\cite{SETI-blog}.

The remainder of this paper is structured as follows. In \cref{sec:background} we present background on object
storage and Apache Spark as well as the motivation for our work. In \cref{sec:algorithm} we describe how Stocator works.
In \cref{sec:methodology} we present the methodology for our performance evaluation, including our experimental set up and a description of our workloads. In \cref{sec:evaluation} we present a detailed evaluation of Stocator, comparing its performance with existing Hadoop object storage connectors, from the point of view of runtime, number of operations and resource utilization. \Cref{sec:related} discusses related work and finally in \cref{sec:conclusion} we conclude.


\section{Background}
\label{sec:background}

We provide background material necessary for understanding the remainder of the paper. First, we describe object storage and then the background on Spark\cite{spark} and its implementation that have implications on the way that it uses object storage. Finally, we motivate the need for Stocator.

\subsection{Cloud Object Storage}
An object encapsulates data and metadata describing the object and its data. An entire object is created at once and cannot be updated in place, although the entire value of an object can be replaced. Object storage is typically accessed through RESTful HTTP, which is a good fit for cloud applications. This simple object semantics enables the implementation of highly scalable, distributed and durable object storage that can provide very large storage capacities at low cost.
Object storage is ideal for storing unstructured data, e.g., video, images, backups and documents such as web pages and blogs. Examples of object storage systems include AWS S3\cite{s3}, Azure Blob storage\cite{azure}, OpenStack Swift\cite{openstack-swift} and IBM Cloud Object Storage\cite{ibm-cloud-object-storage}.

Object storage has a shallow hierarchy. A storage account may contain one or more buckets or containers (hereafter we use the term container), where each container may contain many objects. Typically there is no hierarchy in a container, e.g.,
no containers within a container, although there is support for hierarchical naming. In particular, when listing the contents of a container a separator character, e.g., ``/'' or ``*'', between levels of naming can be specified as well as a prefix string, so that only the names for objects in the container starting with the prefix will be included.
This is different than file systems where there is both hierarchy in the implementation as well as in naming, i.e., a directory is a special file that can contain other files and directories.

Common operations on object storage include:
\begin{enumerate}
	\item \emph{PUT Object}, which creates an object, with the name, data and metadata provided with the operation,
	\item \emph{GET object}, which returns the data and metadata of the object,
	\item \emph{HEAD Object}, which returns just the metadata of the object,
	\item \emph{GET Container}, which lists the objects in a container,
	\item \emph{HEAD Container}, which returns the metadata of a container, and
	\item \emph{DELETE Object}, which deletes an object.
\end{enumerate}
Object creation is atomic, so that two simultaneous PUTs on the same name will create an object with the data of one PUT , but not some combination of the two.

In order to enable a highly distributed implementation the consistency semantics for object storage often include some degree of \emph{eventual consistency}\cite{vogels_2009}. 
Eventual consistency guarantees that if no new updates are made to a given data item, then eventually all accesses to that item will return the same value. There are various aspects of eventual consistency. For example, AWS\cite{aws} guarantees read after write consistency for its S3 object storage system, i.e., that a newly created object will be instantly visible.
Note that this does not necessarily include read after update, i.e., that a new value for an existing object name will be instantly visible, or read after delete, i.e., that a delete will make an object instantly invisible.
Another aspect of eventual consistency concerns the listing of the objects in a container; the creation and deletion of an object may be eventually consistent with respect to the listing of its container. In particular, a container listing may not include a recently created object and may not exclude a recently deleted object.

\subsection{Spark}
We describe Spark's execution model, how Spark interacts with storage, pointing out some of the problems that arise when Spark works on data in object storage.

\subsubsection{Spark execution model}
The execution of a Spark application is orchestrated by the \emph{driver}. The driver divides the application into \emph{jobs} and jobs into \emph{stages}. One stage does not begin execution until the previous stage has completed.
Stages consists of \emph{tasks}, where each task is totally independent of the other tasks in that stage, so that the tasks can be executed in parallel. The output of one stage is typically passed as the input to the next stage, so that a task reads its input from the output of the previous stage and/or from storage. Similarly, a task writes its output to the next stage and/or to storage. The driver creates worker processes called \emph{executors} to which it assigns the execution of the tasks.

The execution of a task may fail. In that case the driver may start a
new execution of the same task. The execution of a task may also be slow and in some cases the driver cannot tell whether the execution has failed or is just slow.
Spark has an important feature to deal with these cases called \emph{speculation}, where it speculatively executes multiple executions of the same task in parallel. Speculation can cut down on the total elapsed time for a Spark
application/job. Thus, a task may be executed multiple times and each such \emph{attempt} to execute a task is assigned a unique identifier, containing a job identifier, a task identifier and an execution attempt number.

\subsubsection{Spark and its underlying storage}
\begin{figure}[t!]
	\includegraphics[width=\columnwidth]{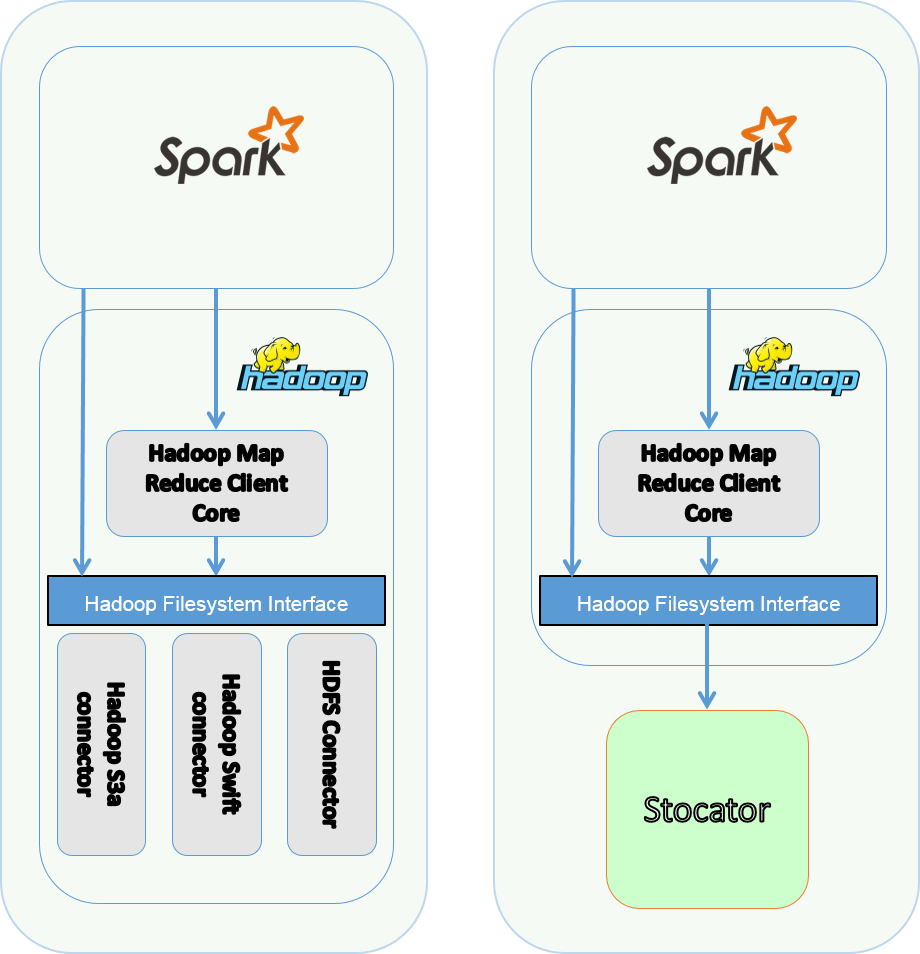}
	\caption{Hadoop Storage Connectors}
	\label{fig:HadoopStorageConnectors}
\end{figure}

Spark interacts with its storage system through Hadoop\cite{hadoop}, primarily through a component called the Hadoop Map Reduce Client Core (HMRCC) as shown in the diagram on the left side in \cref{fig:HadoopStorageConnectors}.

HMRCC interacts with its underlying storage through the Hadoop File System Interface. A connector that implements the interface must be implemented for each underlying storage system. For example, the Hadoop distribution includes a connector for HDFS, as well as an S3a connector for the S3 object store API and a Swift connector for the OpenStack Swift object store API.

A task writes output to storage through the Hadoop \emph{File\-Output\-Committer}. Since each task execution attempt needs to write an output file of the same name, Hadoop employs a rename strategy, where each execution attempt writes its own task temporary file. At \emph{task commit}, the output committer renames the task temporary file to a job temporary file.
Task commit is done by the executors, so it occurs in parallel. And then when all of the tasks of a job complete,
the driver calls the output committer to do \emph{job commit}, which renames the job temporary files to their final names. 
Job commit occurs in the driver after all of the tasks have committed and does not benefit from parallelism.
\begin{figure}[t!]
	\centering
	\includegraphics[width=1.01\columnwidth]{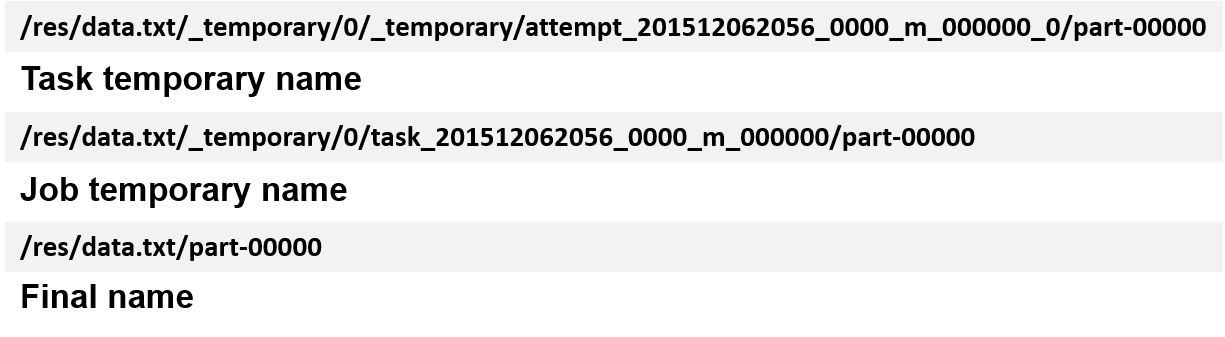}
	\caption{Sequence of names for part 0 of output from task temporary name to job temporary name to final name.}
	\label{fig:SequenceOfNames}
\end{figure}
\Cref{fig:SequenceOfNames} shows the names for a task's output.

This two stage strategy of task commit and then job commit was chosen to avoid the case where incomplete results might be interpreted as complete results. However, Hadoop also writes a zero length object with the name \_SUCCESS when a job completes successfully, so the case of incomplete results can easily by identified by the absence of a \_SUCCESS object.
Accordingly, there is now a new version of the file output committer
algorithm (version 2), where the task temporary files are renamed to their final names at task commit and job commit is largely reduced to the writing of the \_SUCCESS object.
However, as of Hadoop 2.7.3, this algorithm is not yet the default output committer.

Hadoop is highly distributed and thus it keeps its state in its storage system, e.g., HDFS or object storage.
In particular, the output committer determines what temporary objects need to be renamed through directory listings, i.e.,
it lists the directory of the output dataset to find the directories and files holding task temporary and job temporary output. In object stores this is done through container listing operations. However, due to eventual consistency a container listing may not contain an object that was just successfully created, or it may still contain an object that was just successfully deleted. This can lead to situations where some of the legitimate output objects do not get renamed by the output committer, so that the output of the Spark/Hadoop job will be incomplete.

This danger is compounded when speculation is enabled, and thus, despite the benefits of speculation, Spark users are encouraged to run with it disabled. Furthermore, in order to avoid the dangers of eventual consistency entirely,
Spark users are often encouraged to copy their input data to HDFS, run their Spark job over the data in HDFS, and then when it is complete, copy the output from HDFS back to object storage. Note, however, that this adds considerable overhead.
Existing solutions to this problem require a consistent storage system in addition to object storage\cite{s3mper-blog,emrfs-blog,s3guard-issue}.


\subsection{Motivation}
To motivate the need for Stocator we show the sequence of interactions between Spark and its storage system for a
program that executes a single task that produces a single output object as shown in \cref{fig:ProduceOneObject}.
\begin{figure}[]
	\centering
	\begin{lstlisting}[basicstyle=\small,columns=fullflexible]
	val data = Array(1)
	val distData = sc.parallelize(data)
	val finalData = distData.coalesce(1)
	finalData.saveAsTextFile("hdfs://res/data.txt")
	\end{lstlisting}
	\caption{A Spark program that executes a single task that produces a single output object.}
	\label{fig:ProduceOneObject}
\end{figure}
Spark and Hadoop were originally designed to work with a file system. Accordingly, \cref{tab:FileSystemOperations} shows the series of file system operations that Spark carries out for the sample program.
\begin{table*}[]
	\centering
	\renewcommand{\arraystretch}{1.3}
	\resizebox{\textwidth}{!}{%
	\begin{tabular}{|c|l|l|}
		\hline
		\multicolumn{1}{|l|}{} & \multicolumn{1}{c|}{\textbf{Operation}}                                                                        & \multicolumn{1}{c|}{\textbf{File}}                                                          \\ \hline\hline
		1                      & \begin{tabular}[c]{@{}l@{}}Spark Driver: make directories recursively\end{tabular}                           & hdfs://res/data.txt/\_temporary/0                                                                  \\ \hline
		2                      & \begin{tabular}[c]{@{}l@{}}Spark Executor: make directories recursively\end{tabular}                         & hdfs://res/data.txt/\_temporary/0/\_temporary/attempt\_201702221313\_0000\_m\_000001\_1            \\ \hline
		3                      & \begin{tabular}[c]{@{}l@{}}Spark Executor: write task temporary object\end{tabular}                          & hdfs://res/data.txt/\_temporary/0/\_temporary/attempt\_201702221313\_0000\_m\_000001\_1/part-00001 \\ \hline
		4                      & \begin{tabular}[c]{@{}l@{}}Spark Executor: list directory\end{tabular}                                       & hdfs://res/data.txt/\_temporary/0/\_temporary/attempt\_201702221313\_0000\_m\_000001\_1            \\ \hline
		5                      & \begin{tabular}[c]{@{}l@{}}Spark Executor:\\ rename task temporary object to job temporary \\object\end{tabular} & hdfs://res/data.txt/\_temporary/0/task\_201702221313\_0000\_m\_000001/part-00001                   \\ \hline
		6                      & \begin{tabular}[c]{@{}l@{}}Spark Driver:\\ list job temporary directories recursively\end{tabular}             & hdfs://res/data.txt/\_temporary/0/task\_201702221313\_0000\_m\_000001                              \\ \hline
		7                      & \begin{tabular}[c]{@{}l@{}}Spark Driver:\\ rename job temporary object to final name\end{tabular}              & hdfs://res/data.txt/part-00001                                                                     \\ \hline
		8                      & Spark Driver: write \_SUCCESS object                                                                           & hdfs://res/data.txt/\_SUCCESS                                                                   \\ \hline
	\end{tabular}%
	}
\caption{The file system operations executed on behalf of a Spark program that executes a single task to produces a single output object.}
\label{tab:FileSystemOperations}
\end{table*}
\begin{enumerate}
	\item The Spark driver and executor recursively create the directories for the task temporary, job temporary and final output (steps 1--2).
	\item The task outputs the task temporary file (step 3).
	\item At task commit the executor lists the task temporary directory, and renames the file it finds to its job temporary name (steps 4-5).
	\item At job commit the driver recursively lists the job temporary directories and renames the file it finds to its final names (steps 6-7).
	\item The driver writes the \_SUCCESS object.
\end{enumerate}

When this same Spark program runs with the Hadoop Swift or S3a connectors, these file operations are translated to equivalent operations on objects in the object store. These connectors use PUT to create zero byte objects representing the
directories, after first using HEAD to check if objects for the directories already exist. When listing the contents of a directory, these connectors  descend the ``directory tree'' listing each directory. To rename objects these connectors use PUT or COPY to copy the object to its new name and then use DELETE on the object at the old name. All of the zero byte directory objects also need to be deleted. Overall the Hadoop Swift connector executes 48 REST operations and the S3a connector executes 117 operations. \Cref{tab:RESTOperations} shows the breakdown according to operation type.
\begin{table}[]
	\centering
	\renewcommand{\arraystretch}{1.3}
	\resizebox{\columnwidth}{!}{%
		\begin{tabular}{|c|c|c|c|c|c|c|}
			\hline 
			& \thead{HEAD \\ Object} & \thead{PUT \\ Object} & \thead{COPY \\ Object} & \thead{DELETE \\ Object} & \thead{GET \\ Cont.} & Total \\ 
			\hline 
			Hadoop-Swift 	& $25$ 	& $7$  	& $3$ & $8$ & $5$ 	& $48$ \\ 
			\hline
			S3a 			& $71$ 	& $5$ 	& $2$ & $4$ & $35$ 	& $117$ \\ 
			\hline
			Stocator 		& $4$ 	& $3$ 	& $-$ & $-$ & $1$ 	& $8$ \\
			\hline
		\end{tabular}
	} 
	\caption{Breakdown of REST operations by type for the Spark program that creates an output consisting of a single object.}
	\label{tab:RESTOperations}
\end{table}

In the next section we describe Stocator,
which leverages object storage semantics to replace the temporary
file/rename paradigm and takes advantage of hierarchal naming to avoid
the creation of directory objects.
For the Spark program in \cref{fig:ProduceOneObject} Stocator executes just 8
REST operations: 3 PUT object, 4 HEAD object and 1 GET container.


\section{Stocator algorithm}
\label{sec:algorithm}

The right side of \cref{fig:HadoopStorageConnectors} shows how
Stocator fits underneath HMRCC; it implements the Hadoop Filesystem
Interface just like the other storage connectors.
Below we describe the basic Stocator protocol;
and then how it streams data, deals with eventual consistency, and
reduces operations on the read path. 
Finally we provide several examples of the protocol in action.

\subsection{Basic Stocator protocol}
\label{subsec:basicStocatorProtocol}

The overall strategy used by Stocator to avoid rename is to write
output objects directly to their final name
and then to determine which objects
actually belong to the output at the time that the output is read by
its consumer,
e.g., the next Spark job in a sequence of jobs.
Stocator does this in a way that preserves the fault tolerance model
of Spark/Hadoop and enables speculation.
Below we describe the components of this strategy.

As described in \cref{sec:background} the driver orchestrates
the execution of a Spark application.
In particular,
the driver is responsible for creating a ``directory'' to hold an
application's output dataset.
Stocator uses this ``directory'' as a  marker to indicate that it wrote the
output.
In particular, Stocator writes a zero byte object with the name of the
dataset
and object metadata that indicates that the object was written by
Stocator.
All of the
dataset's parts are stored hierarchically under this name.

Then when a Spark task asks to create a temporary object for its part
through HMRCC, Stocator recognizes the pattern of the name and writes
the object directly to its final name so it will not need to be
renamed. If Spark executes a task multiple times due to failures, slow
execution or speculative execution,
each execution attempt is assigned a number.
The Stocator object naming scheme includes this attempt number so that
individual attempts can be distinguished.
In particular, HMRCC asks to write a temporary
file/object in a temporary directory of the form 
\url{<output-dataset-name>/_temporary/0/_temporary/attempt_<job-timestamp>_0000_m_000000_<attempt-number>/part-<part-number>},
where \url{<job-timestamp>} is the timestamp of the Spark job,
\url{<attempt-number>} is the number of attempt, and
\url{<part-number>} is the number of the part.
Stocator notices this pattern and
in place of the temporary object in the temporary directory,
it writes an object with the name
\url{<output-dataset-name>/part-<part-number>_attempt_<job-timestamp>_0000_m_000000_<attempt-number>}.

Finally, when all tasks have completed successfully, Spark writes a
\_SUCCESS object through HMRCC.
Notice that by avoiding rename, Stocator also avoids the need for list
operations during task and job commit that may lead to incorrect
results due to eventual consistency;
thus, the presence of a \_SUCCESS object means that there was a
correct execution for each task and that there is an object for
each part in the output.

\subsection{Alternatives for reading an input dataset}
Stocator delays the determination of which parts belong to an output
dataset until it reads the dataset as input.
We consider two options.

The first option is simpler to implement since it can be done entirely
in the implementation of Stocator.
It depends on the assumption that Spark exhibits fail-stop behavior, 
i.e., that a Spark server executes correctly until it halts.
After determining that the dataset was produced by Stocator through
reading the metadata from the object written with the dataset's name,
and checking that the \_SUCCESS object exists,
Stocator lists the object parts belonging to the dataset through a GET
container operation.
If there are objects in the list representing multiple execution
attempts for same task,
Stocator will choose the one that has the most data. 
Given the fail-stop assumption,
the fact that all successful execution attempts write the same output,
and that it is certain that at least one attempt succeeded (otherwise there
would not be a \_SUCCESS object),
this is the correct choice.

The second option is more complex to implement.
Here at the time the \_SUCCESS object is written, Stocator includes in
it a list of all the successful execution attempts completed by the
Spark job.
Now after determining that the dataset was produced by Stocator through
reading the metadata from the object written with the dataset's name,
and checking that the \_SUCCESS object exists,
Stocator reads the manifest of successful task execution attempts from
the \_SUCCESS object.
Stocator uses the manifest to reconstruct the list of constituent object parts 
of the dataset.
In particular, the construction of the object part names follows the
same pattern outlined above that was used when the parts were
written. 

The benefit of the second option is that it solves the remaining
eventual consistency issue by constructing the object names from the
manifest rather than issuing a REST command to list the object
parts, which may not return a correct result in the presence of
eventual consistency.
The second option also does not need the fail-stop assumption.
However, due to its simplicity we have implemented the first option in
our Stocator prototype.

\subsection{Streaming of output}
When Stocator outputs data it streams the data to the object store as
the data is produced using chunked transfer encoding.
Normally the total length of the object is one of the parameters of a
PUT operation and thus needs to be known before starting the
operation. Since Spark produces the data for an object on the fly and
the final length of the data is not known until all of its data is
produced, this would mean that Spark would need to store the entire
object data 
prior to starting the PUT. To avoid running out of memory, a storage
connector for Spark can store the object in the Spark server's local
file system as the connector produces the object's content, and then
read the object back from the file to do the PUT operation on the
object store. Indeed this is what the default Hadoop Swift and S3a
connectors do.  
Instead Stocator leverages HTTP chunked transfer encoding, which is
supported by the Swift API. In chunked transfer encoding the object
data is sent in chunks, the sender needs to know the length of each
chunk, but it does not need to know the final length of the object
content before starting the PUT operation. 
S3a has an optional feature, not activated by default,
called fast upload, where it leverages the multi-part upload feature
of the S3 
API. This achieves a similar effect to chunked transfer encoding
except that it uses more memory since the minimum part size 
for multi-part upload is 5 MB.

\begin{table*}[ht]
	\centering
	\renewcommand{\arraystretch}{1.3}
	\resizebox{\textwidth}{!}{%
		\begin{tabular}{|l|l|l|}
			\hline
			\multicolumn{1}{|c|}{\textbf{}} & \multicolumn{1}{c|}{\textbf{Hadoop Map Reduce Client Core}} & \multicolumn{1}{c|}{\textbf{Stocator}} \\ \hline\hline
			1 & \begin{tabular}[c]{@{}l@{}}PUT /res/data.txt/\_temporary/0/\_temporary/attempt\_201512062056\_0000\_m\_000000\_0/part-00000\end{tabular} & PUT /res/data.txt/part-00000\_attempt\_201512062056\_0000\_m\_000000\_0 \\ \hline
			2 & PUT /res/data.txt/\_temporary/0/\_temporary/attempt\_201512062056\_0000\_m\_000000\_0/part-00001 & PUT /res/data.txt/part-00001\_attempt\_201512062056\_0000\_m\_000000\_0 \\ \hline
			3 & PUT /res/data.txt/\_temporary/0/\_temporary/attempt\_201512062056\_0000\_m\_000000\_0/part-00002 & \begin{tabular}[c]{@{}l@{}}PUT /res/data.txt/part-00002\_attempt\_201512062056\_0000\_m\_000000\_0\end{tabular} \\ \hline
			4 & PUT /res/data.txt/\_temporary/0/\_temporary/attempt\_201512062056\_0000\_m\_000000\_1/part-00002 & \begin{tabular}[c]{@{}l@{}}PUT /res/data.txt/part-00002\_attempt\_201512062056\_0000\_m\_000000\_1\end{tabular} \\ \hline
			5 & PUT /res/data.txt/\_temporary/0/\_temporary/attempt\_201512062056\_0000\_m\_000000\_2/part-00002 & \begin{tabular}[c]{@{}l@{}}PUT /res/data.txt/part-00002\_attempt\_201512062056\_0000\_m\_000000\_2\end{tabular} \\ \hline
			6 & DELETE /res/data.txt/\_temporary/0/\_temporary/attempt\_201512062056\_0000\_m\_000000\_0/part-00002 & \begin{tabular}[c]{@{}l@{}}DELETE /res/data.txt/part-00002\_attempt\_201512062056\_0000\_m\_000000\_0\end{tabular} \\ \hline
			7 & DELETE /res/data.txt/\_temporary/0/\_temporary/attempt\_201512062056\_0000\_m\_000000\_2/part-00002 & \begin{tabular}[c]{@{}l@{}}DELETE /res/data.txt/part-00002\_attempt\_201512062056\_0000\_m\_000000\_2\end{tabular} \\ \hline
			8 & \begin{tabular}[c]{@{}l@{}}Task commits and job commit generate 2 pairs of COPY and DELETE for each successful attempt\end{tabular} & No operations are performed here \\ \hline
			9 & PUT /res/data.txt/\_SUCCESS & PUT /res/data.txt/\_SUCCESS \\ \hline
		\end{tabular}%
	}
	\caption{Possible operations performed by the Spark application showed in \cref{fig:ProduceThreeObject}}
	\label{tab:possibleOperations}
\end{table*}

\subsection{Optimizing the read path}
We describe several optimizations that Stocator uses to reduce the
number of operations on the read path.

The first optimization can remove a HEAD operation that occurs just
before a GET operation for the same object.
In particular, the storage connector often reads the metadata of an object
just before its data.
Typically this is to check that the object exists and to obtain the
size of the object.
In file systems this is performed by two different operations.
Accordingly a naive implementation for object storage would read object
metadata through a HEAD operation, 
and then read the data of the object itself through a GET operation.
However, object store GET operations also return the metadata of an
object together with its data.
In many of these cases Stocator is able to remove the HEAD operation,
which can greatly reduce the overall number of operations invoked on
the underlying object storage system. 

A second optimization is caching the results of HEAD operations.
A basic assumption of Spark is that the input is immutable.
Thus, if a HEAD is called on the same object multiple times,
it should return the same result.
Stocator uses a small cache to reduce these calls.

\subsection{Examples}

\begin{figure}[]
	\centering
	\begin{lstlisting}[basicstyle=\small,columns=fullflexible]	
	val data = Array(1, 2, 3)
	val distData = sc.parallelize(data)
	distData.saveAsTextFile("swift2d://res.sl/data.txt")
	\end{lstlisting}
	\caption{A Spark program where three tasks each write an object part.}
	\label{fig:ProduceThreeObject}
\end{figure}

We show here some examples of Stocator at work.
For simplicity we focus on Stocator's interaction with HMRCC to
eliminate the rename paradigm and so we do not show all of the requests
that HMRCC makes on Stocator, e.g.,
to create/delete ``directories'' and check their status.

\Cref{fig:ProduceThreeObject} shows a simple Spark program
that will be executed by three tasks, each task writing its part to
the output dataset called $data.txt$ in a container called $res$.
The {\it swift2d:} prefix in the URI for the output dataset indicates that
Stocator is to be used as the storage connector.
\Cref{tab:possibleOperations} shows the operations that can be executed by our example in different situations.

Lines 1-3 and 8-9 are executed when each task
runs exactly once and the program completes successfully. We show the
requests that HMRCC generates; for each task it issues one request to create a
temporary object and two requests to ``rename'' it (copy to a new name
and delete the object at the former name). We see that Stocator
intercepts the pattern for the temporary name that it receives from
HMRCC, and creates the final names for the objects directly. At the
end of the run Spark creates the \_SUCCESS object.

Lines 1-5, instead, shows an execution where Spark
decides to execute Task 2 three times, i.e., three attempts.
This could be because the first and second attempts failed
or due to speculation because they were slow. Notice that Stocator includes the
attempt number as part of the name of the objects that it creates.

By adding lines 6-9 to the previous, we show what happens when Spark is able to clean up the results from the
duplicate attempts to execute Task 2.
In particular, Spark aborts attempts 0 and 2, and commits attempt
1. When Spark aborts attempts 0 and 2, HMRCC deletes their
corresponding temporary objects. Stocator recognizes the pattern for
the temporary objects and deletes the corresponding objects that it
created.

If Spark is
not able to clean up the results from the duplicate attempts to
execute Task 2, we have lines 1-5 and 8-9.
In particular, we see that Stocator created five object parts, one
each for Tasks 0 and 1, and three for Task 2 due to its extra
attempts. We assume as in the previous situation that it is attempt 1
for Task 2 that succeeded.
Stocator recognizes this through the
manifest stored in the \_SUCCESS object.

\section{Methodology}
\label{sec:methodology}

We describe the experimental platform, deployment scenarios, workloads and performance metrics that we use to evaluate Stocator.

\subsection{Experimental Platform}
\label{subsec:exp_plat}


Our experimental infrastructure includes a Spark cluster, an IBM Cloud
Object Storage (formerly Cleversafe) cluster, Keystone, and
Graphite/Grafana. The Spark cluster consists of three bare metal
servers. 
Each server has a dual Intel Xeon E52690 processor with 12
hyper-threaded 2.60 GHz cores (so 24 hyper-threaded cores per server),
256 GB memory, a 10 Gbps NIC and a 1 TB SATA disk.
That means that the total parallelism of the Spark
cluster is 144. We run 12 executors on each server; each executor gets
4 cores and 16 GB of memory. We use Spark submit to run the
workloads and the driver runs on one of the Spark servers (always the
same server). We use the standalone Spark cluster manager.

Our IBM Cloud Object Storage (COS) \cite{resch_2011} cluster also runs on bare metal. It consists of two Accessers,
front end servers that receive the REST commands and then orchestrate their execution across twelve Slicestors, which hold the storage. Each Accesser has two 10 Gbps NICs bonded to yield 20 Gbps. Each Slicestor has twelve 1 TB SATA disks for data.
The Information Dispersal Algorithm (IDA) or erasure code is (12, 8, 10), which means that the erasure code splits the data into 12 parts, 8 parts are needed to read the data, and at least 10 parts need to be written for a write to complete. IBM COS exposes multiple object APIs; we use the Swift and S3 APIs. 

We employ HAProxy for load balancing. It is installed on each of the Spark servers and configured with round-robin so that connections opened by a Spark server with the object storage alternate between Accessers. Given that each of the three Spark servers has a 10 Gbps NIC, the maximum network bandwidth between the Spark cluster and the COS cluster is 30 Gbps.

Keystone and Graphite/Grafana run on virtual machines. Keystone provides authentication/authorization for the Swift API. We collect monitoring data on Graphite and view it through Grafana to check that there are no unexpected bottlenecks during the performance runs. In particular we use the Spark monitoring interface and the collectd daemon to collect monitoring data from the Spark servers, and we use the Device API of IBM COS to collect monitoring data from the Accessers and the Slicestors.

\subsection{Deployment scenarios}
\label{subsec:scenarios}


In our experiments, we compare Stocator with the Hadoop Swift and S3a
connectors.
By using different configurations of these two connectors, we define
six scenarios: (i)~Hadoop-Swift Base (\textbf{H-S Base}),
(ii)~S3a Base (\textbf{S3a Base}), (iii)~Stocator Base
(\textbf{Stocator}), (iv)~Hadoop-Swift Commit V2 (\textbf{H-S Cv2}),
(v)~S3a Commit V2 (\textbf{S3a Cv2}) and
(vi)~S3a Commit V2 + Fast Upload (\textbf{S3a Cv2+FU}).
These scenarios are split into 3 groups according to the optional optimization
features that are active.
The first group, with the suffix \textit{Base},
uses connectors out of the box, meaning that no optional features
are active.
The second group, with the suffix \textit{Commit V2}, uses
the version 2 of Hadoop FileOutputCommitter that
reduces the number of copy operations towards the object storage
(as described in Section~\ref{sec:background}).
The last group, with the
suffix \textit{Commit V2 + Fast Upload}, uses both version 2 of
Hadoop FileOutputCommitter and an optimization feature of
S3a called S3AFastOutputStream that streams data to the object storage as it is produced (as described in Section~\ref{sec:algorithm}). 

All experiments run on \textit{Spark 2.0.1} with a
patched~\cite{hadoop-aws-jiira} version of Hadoop 2.7.3
infrastructure.
This
patch allows us to use, for the S3a scenarios, \textit{Amazon SDK
  version 1.11.53} instead of version 1.7.4.
The Hadoop-Swift
scenarios run with the default Hadoop-Swift connector that comes with
Hadoop 2.7.3. Finally, the Stocator scenario runs with
\textit{stocator~1.0.8}. 
 

\subsection{Benchmark and Workloads}
\label{subsec:benchmark}

\begin{table}[!t]
	\centering
	\renewcommand{\arraystretch}{1.3}
	\resizebox{0.6\columnwidth}{!}{%
		\begin{tabular}{c|c|c|}
			Workload		& Input Size		& Output Size \\ \hline
			Read-Only		& 46.5 GB			&  0 MB \\
			Read-Only 10x	& 465.6 GB			&  0 MB \\
			Teragen			& 0 GB				&  46.5 GB \\
			Copy			& 46.5 GB			&  46.5 GB \\
			Wordcount		& 46.5 GB			&  1.28 MB \\
			Terasort		& 46.6 GB			&  46.5 GB \\
			TPC-DS			& 13.8 GB			&  0 MB \\	
		\end{tabular}
	}
	\caption{Workloads' details.}
	\label{tab:workloads-details}	
\end{table}

To study the performance of our solution we use several workloads
(described in \cref{tab:workloads-details}), that are currently used
in popular benchmark suites and cover different kinds of
applications. The workloads span from simple applications that target a
single and specific feature of the connectors (micro benchmarks), to
real complex applications composed by several jobs (macro
benchmarks).

The micro benchmarks use three different applications: (i)~Read-only,
(ii)~Write-only and (iii)~Copy. The Read-only application reads two
different text datasets, one whose size is 46.5 GB and the second
465.6 GB, and counts the number of lines in them.
For
the Write-only application we use the popular Teragen application,
available in the Spark example suite, that only performs write
operations creating a dataset of 46.5 GB. The last application that we
use for our micro benchmark set is what we call the Copy application;
it copies the small dataset used by the Read-only application.

We also use three macro benchmarks.
The first, Wordcount from Intel Hi-Bench~\cite{hibench, hibench_2011}
test suite, is the ``Hello World'' application for parallel computing.
It is a read-intensive workload, that reads an input text file,
computes the number of times each word occurs in the file and then writes
a much smaller output file containing the word counts.
The second macro benchmark, Terasort, is a popular application used to
understand the performance of large scale computing frameworks like
Spark and Hadoop.
Its input dataset is the output of the Teragen application used in the
micro benchmarks.
The third macro benchmark,TPC-DS, is the Transaction Processing
Performance Council's decision-support benchmark
test \cite{tpcds,tpcds_2006} implemented with DataBricks'
Spark-Sql-Perf library~\cite{databricks_spark_sql_perf}.
It executes several complex queries on files stored in
Parquet format~\cite{parquet}; the input dataset size is 50 GB, which is
compressed to 13.8 GB when converted to Parquet.
The query set that we use to
perform our experiments is composed of the following 8 TPC-DS queries:
q34, q43, q46, q59, q68, q73, q79 and ss\_max.
These are the queries from the \textit{Impala} subset that work
with the Hadoop-Swift connector.
Stocator and S3a support all of the queries in the Impala subset.

The inputs for the Read-only, Copy, Wordcount and Terasort benchmarks
are divided into 128 MB objects.
The outputs of the Copy, Teragen and Terasort benchmarks are also
divided into 128 MB objects.
We also run Spark with a partition size of 128 MB.

\subsection{Performance metrics}
\label{subsec:metrics}
We evaluate the different connectors and scenarios by using
metrics that target the various optimization features. As a general
metric we use the total runtime of the application; this provides a
quick overview of the performance of a specific scenario. To delve
into the reason behind the performance we use two additional metrics. The first
is the number of REST calls -- and their type;
with this metric we are able to understand the load on the
object storage imposed by the connector.
The second metric is the number of bytes read from, written to and
copied in the object storage;
this also help us to understand the load on the object storage imposed
by the connectors.

\section{Experimental Evaluation}
\label{sec:evaluation}

We now present a comparative analysis between the different
scenarios that we defined in~\cref{subsec:scenarios}. We first show
the benefit of Stocator through the average run time of the different
workloads.
Then we compare the number of REST operations issued by the Compute
Layer toward the Object Storage and the relative cost for
these operations charged by cloud object store services.
Finally we compare the number of bytes transferred between the Compute
Layer and the Object Storage.

\subsection{Reduction in run time}
%
%
%
\begin{table*}[]
	\centering
	\renewcommand{\arraystretch}{1.3}
	\resizebox{\textwidth}{!}{%
		\begin{tabular}{|c|c|c|c|c|c|c|c|}
			\hline 
			& Read-Only 50GB & Read-Only 500GB & Teragen & Copy & Wordcount & Terasort & TPC-DS \\ 
			\hline 
			Hadoop-Swift Base & $37.80 \pm 0.48$ & $393.10 \pm 0.92$  & $624.60 \pm 4.00$ & $622.10 \pm 13.52$ & $244.10 \pm 17.72$ & $681.90 \pm 6.10$ & $\textbf{101.50} \pm \textbf{1.50}$ \\ 
			\hline
			S3a Base & $\textbf{33.30} \pm \textbf{0.42}$ & $254.80 \pm 4.00$ & $699.50 \pm 8.40$ & $705.10 \pm 8.50$ & $193.50 \pm 1.80$ & $746.00 \pm 7.20$ & $104.50 \pm 2.20$ \\ 
			\hline
			Stocator & $34.60 \pm 0.56$ & $\textbf{254.10} \pm \textbf{5.12}$ & $\textbf{38.80} \pm \textbf{1.40}$ & $\textbf{68.20} \pm \textbf{0.80}$ & $\textbf{106.60} \pm \textbf{1.40}$ & $\textbf{84.20} \pm \textbf{2.04}$ & $111.40 \pm 1.68$ \\ 
			\hline
			Hadoop-Swift Cv2 & $37.10 \pm 0.54$ & $395.00 \pm 0.80$ & $171.30 \pm 6.36$ & $175.20 \pm 6.40$ & $166.90 \pm 2.06$ & $222.70 \pm 7.30$ & $102.30 \pm 1.16$ \\ 
			\hline  
			S3a Cv2 & $35.30 \pm 0.70$ & $255.10 \pm 5.52$ & $169.70 \pm 4.64$ & $185.40 \pm 7.00$ & $111.90 \pm 2.08$ & $221.90 \pm 6.66$ & $104.00 \pm 2.20$ \\ 
			\hline 
			S3a Cv2 + FU & $35.20 \pm 0.48$ & $\textbf{254.20} \pm \textbf{5.04}$ & $56.80 \pm 1.04$ & $86.50 \pm 1.00$ & $112.00 \pm 2.40$ & $105.20 \pm 3.28$ & $103.10 \pm 2.14$ \\ 
			\hline  
		\end{tabular}
	} 
	\caption{Average run time}
	\label{tab:runtime-comparison}
\end{table*}

\begin{table*}[]
	\centering
	\renewcommand{\arraystretch}{1.3}
	\resizebox{\textwidth}{!}{%
		\begin{tabular}{|c|c|c|c|c|c|c|c|}
			\hline 
			& Read-Only 50GB & Read-Only 500GB & Teragen & Copy & Wordcount & Terasort & TPC-DS \\ 
			\hline 
			Hadoop-Swift Base 	& x$1.09$ & x$1.55$  & x$16.09$ & x$9.12$ & x$2.29$ & x$8.10$ & x$0.91$ \\ 
			\hline
			S3a Base 			& x$0.96$ & x$1.00$ & x$18.03$ & x$10.33$ & x$1.82$ & x$8.86$ & x$0.94$ \\ 
			\hline
			Stocator 			& x$1$ 	& x$1$ & x$1$ & x$1$ & x$1$ & x$1$ & x$1$ \\ 
			\hline
			Hadoop-Swift Cv2 	& x$1.07$ & x$1.55$ & x$4.41$ & x$2.57$ & x$1.57$ & x$2.64$ & x$0.92$ \\ 
			\hline  
			S3a Cv2 			& x$1.02$ & x$1.00$ & x$4.37$ & x$2.72$ & x$1.05$ & x$2.64$ & x$0.93$ \\ 
			\hline 
			S3a Cv2 + FU 		& x$1.02$ & x$1.00$ & x$1.46$ & x$1.27$ & x$1.05$ & x$1.25$ & x$0.93$ \\ 
			\hline  
		\end{tabular}
	} 
	\caption{Workload speedups when using Stocator}
	\label{tab:speedups}
\end{table*}

For each workload we ran each scenario ten times.
We report the average and standard deviation
in~\Cref{tab:runtime-comparison}.
The results shows that, when using a connector out
of the box and under workloads that perform write operations, Stocator
performs much better than Hadoop-Swift and S3a. Only by
activating and configuring optimization features provided by the
Hadoop ecosystem, Hadoop-Swift and S3a manage to close the gap with
Stocator, but they still fall behind. 

\Cref{tab:speedups} shows the speedups that we obtain when using
Stocator with respect to the other connectors. We see a relationship
between Stocator performance and the workload; the more write
operations performed, the greater the benefit obtained. On the one hand the
write-only workloads, like Teragen, run 18 time faster with Stocator
compared to the other out of the box connectors, 4 time faster when
we enable FileOutputCommitter Version 2, and 1.5 times faster when we
also add the S3AFastOutputStream feature. On
the other hand, workloads more skewed toward read operations, like
Wordcount, have lower speedups.

These results are possible thanks to the algorithm implemented in
Stocator. Unlike the alternatives, Stocator removes the rename -- and thus
copy -- operations completely.
In contrast, the other connectors, even with
FileOutputCommitter Version 2,
must still rename each output object once,
although the overhead of the remaining renames is partially masked
since they are carried out by the executors in parallel.

Stocator performs slightly worse
than S3a on two of the workloads that contain only read operations (no writes),
Read-only 50 GB and TPC-DS,
and virtually the same for the larger 500 GB Read-only workload.
We have identified a small start-up cost that we have not yet
removed from Stocator that can explain the difference between the
results for the 50 GB and 500 GB Read-only workload.
As expected the results for the read-only workloads for S3a and Hadoop-Swift
connectors are virtually the same with and without the FileOutputCommitter Version 2
and S3AFastOutputStream features;
these features optimize the write path and do not affect the read path. 

\subsection{Reduction in the number of REST calls}
\begin{figure}[t!]
	\centering
	\includegraphics[width=0.49\columnwidth]{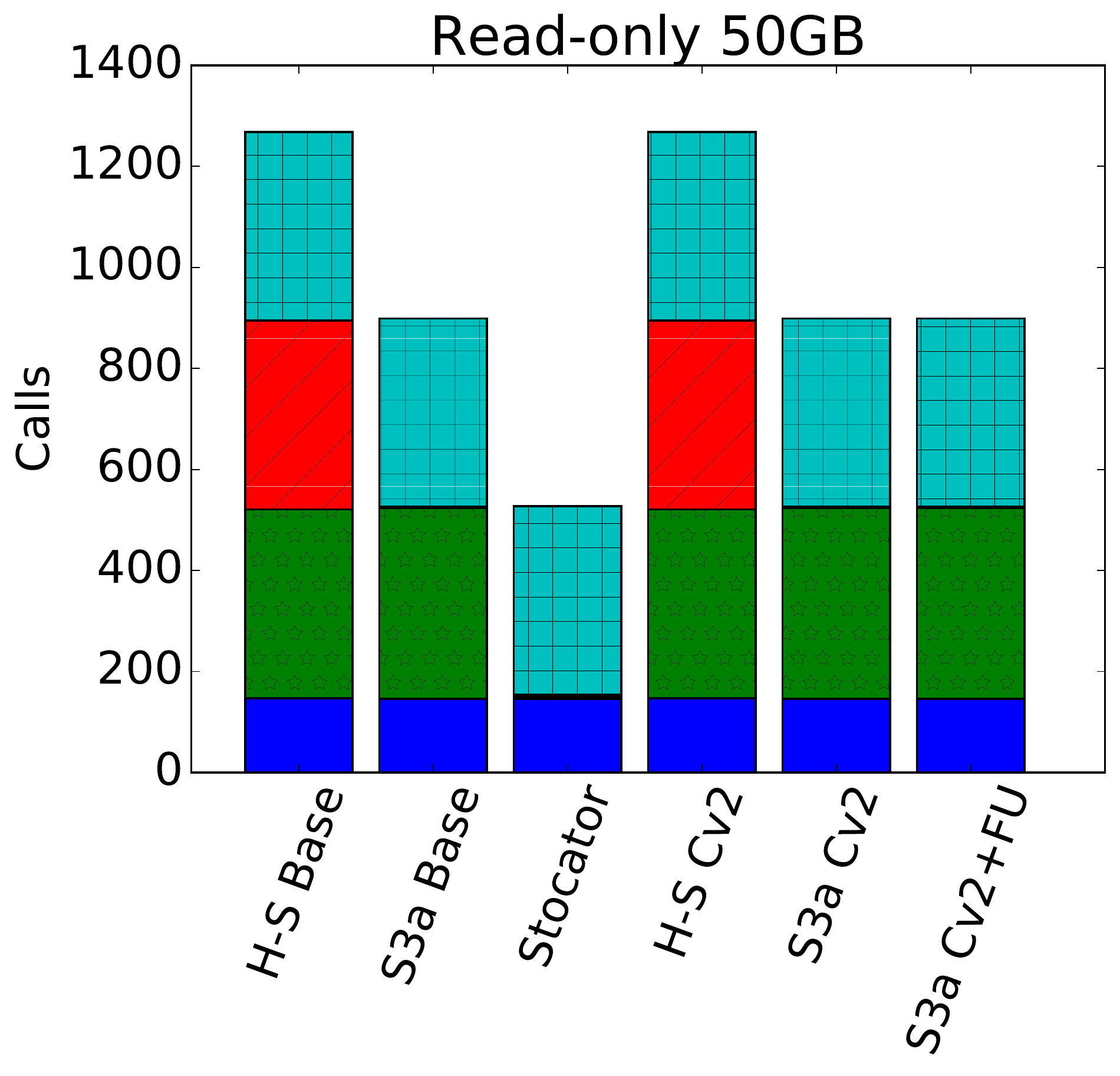}%
	~
	\includegraphics[width=0.49\columnwidth]{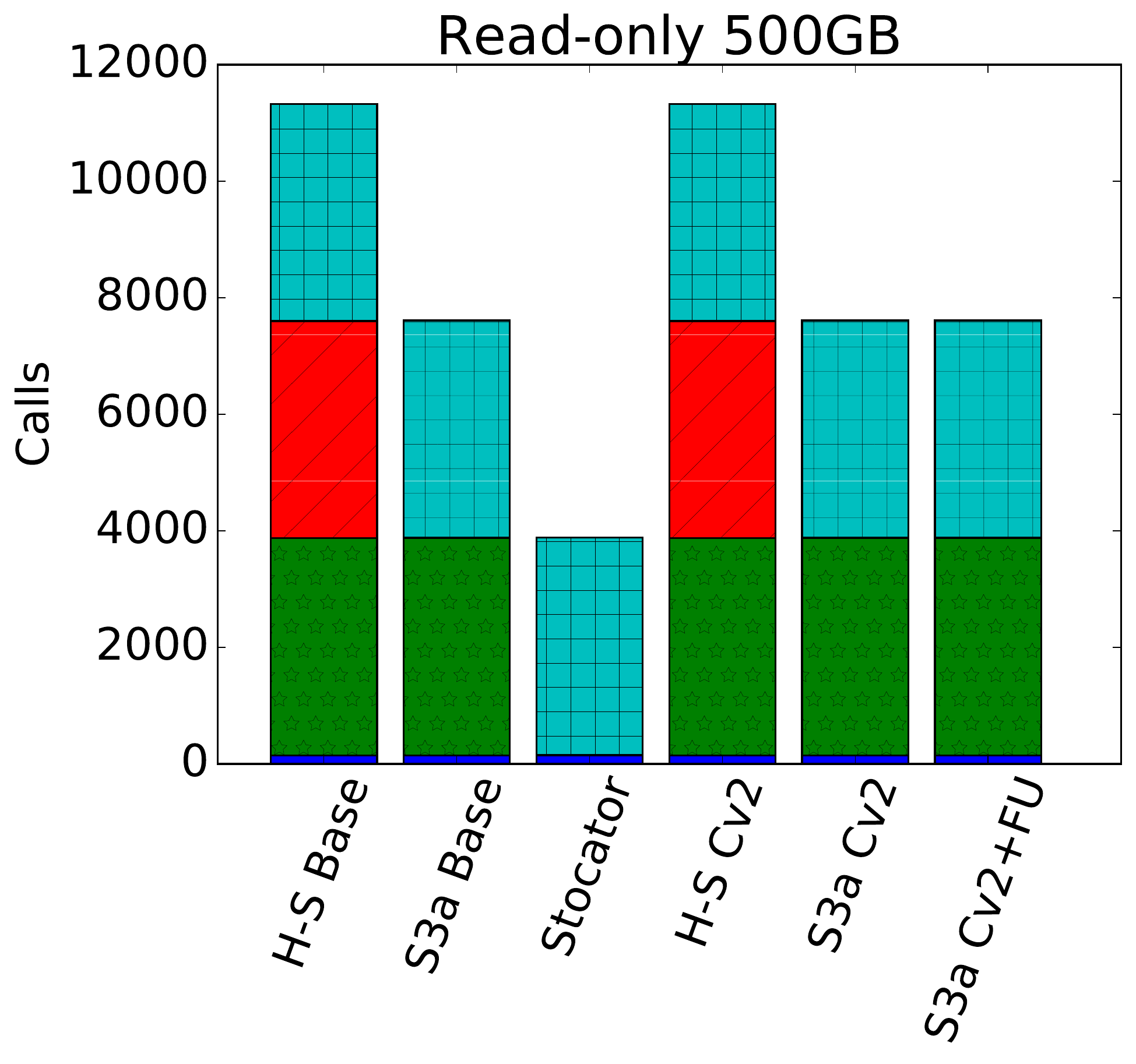}%
	\newline
	\includegraphics[width=0.49\columnwidth]{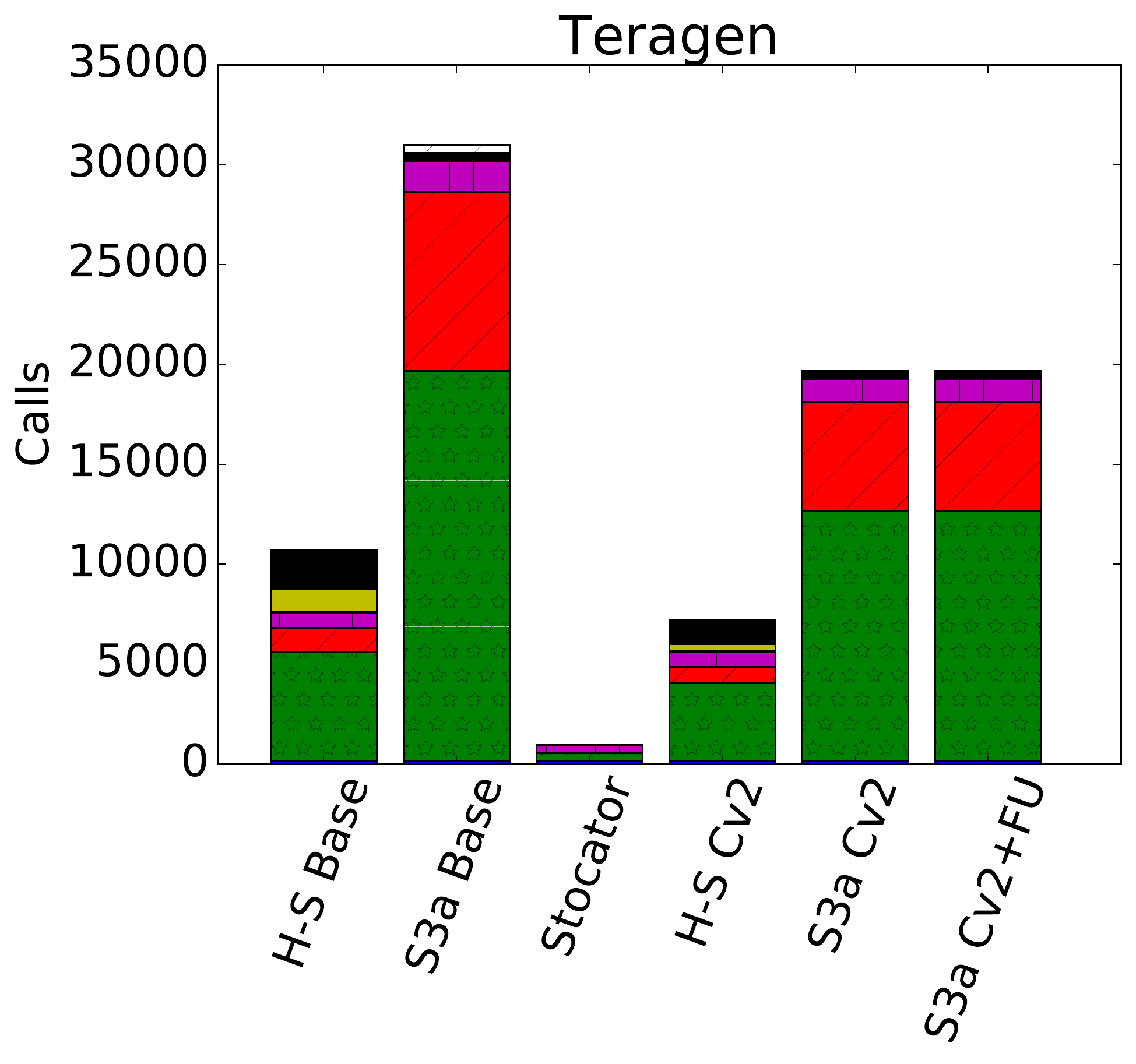}%
	~
	\includegraphics[width=0.49\columnwidth]{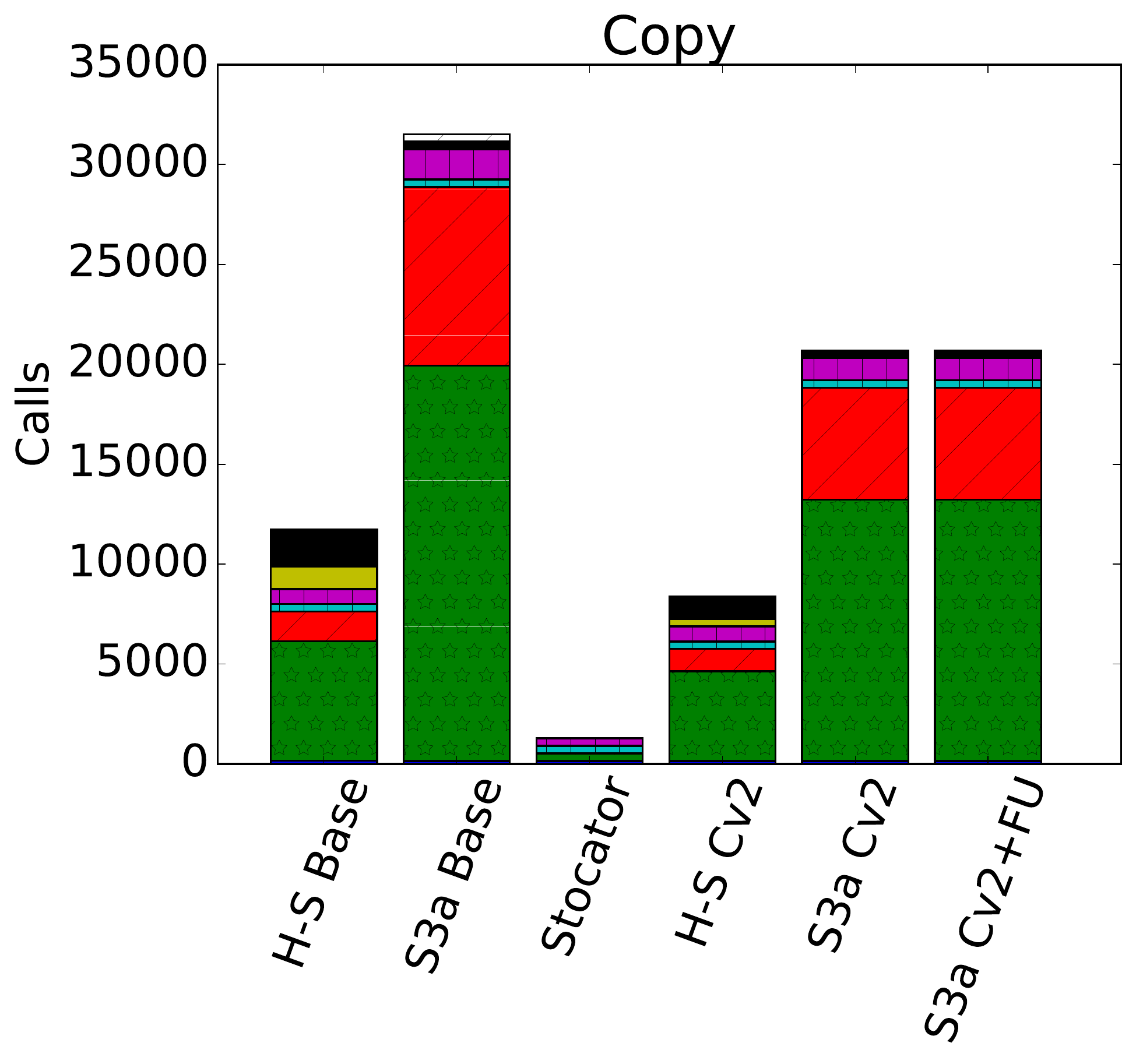}%
	\newline
	\includegraphics[width=\columnwidth]{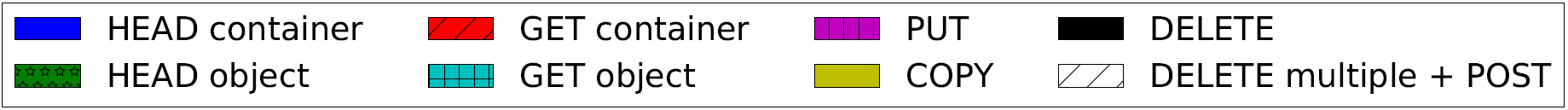}%

	\caption{Micro-benchmarks REST calls comparison}
	\label{fig:micro-rest-comparison}
\end{figure}

\begin{figure*}[t!]
	\centering
	\includegraphics[width=0.3\textwidth]{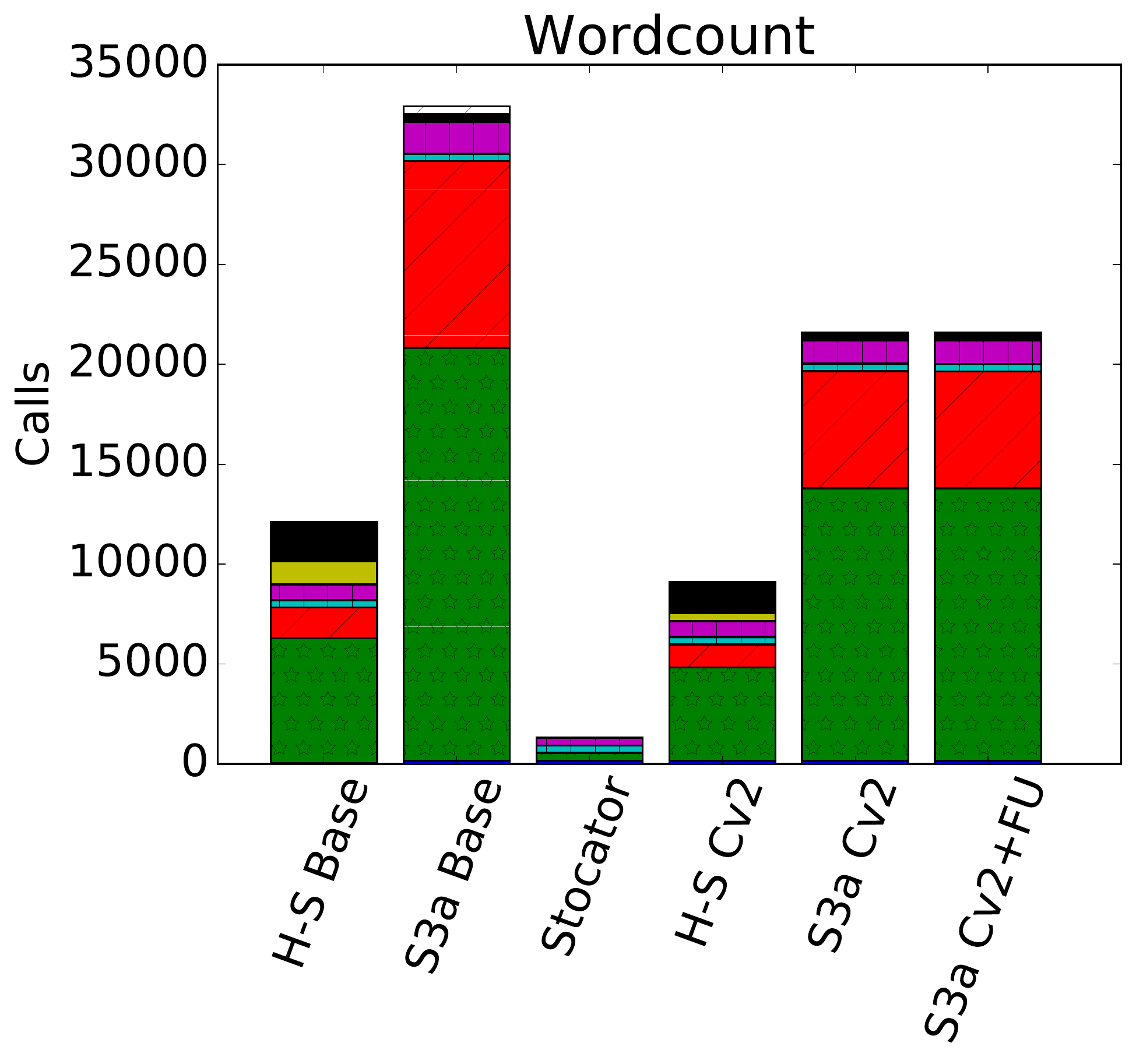}%
	~
	\includegraphics[width=0.3\textwidth]{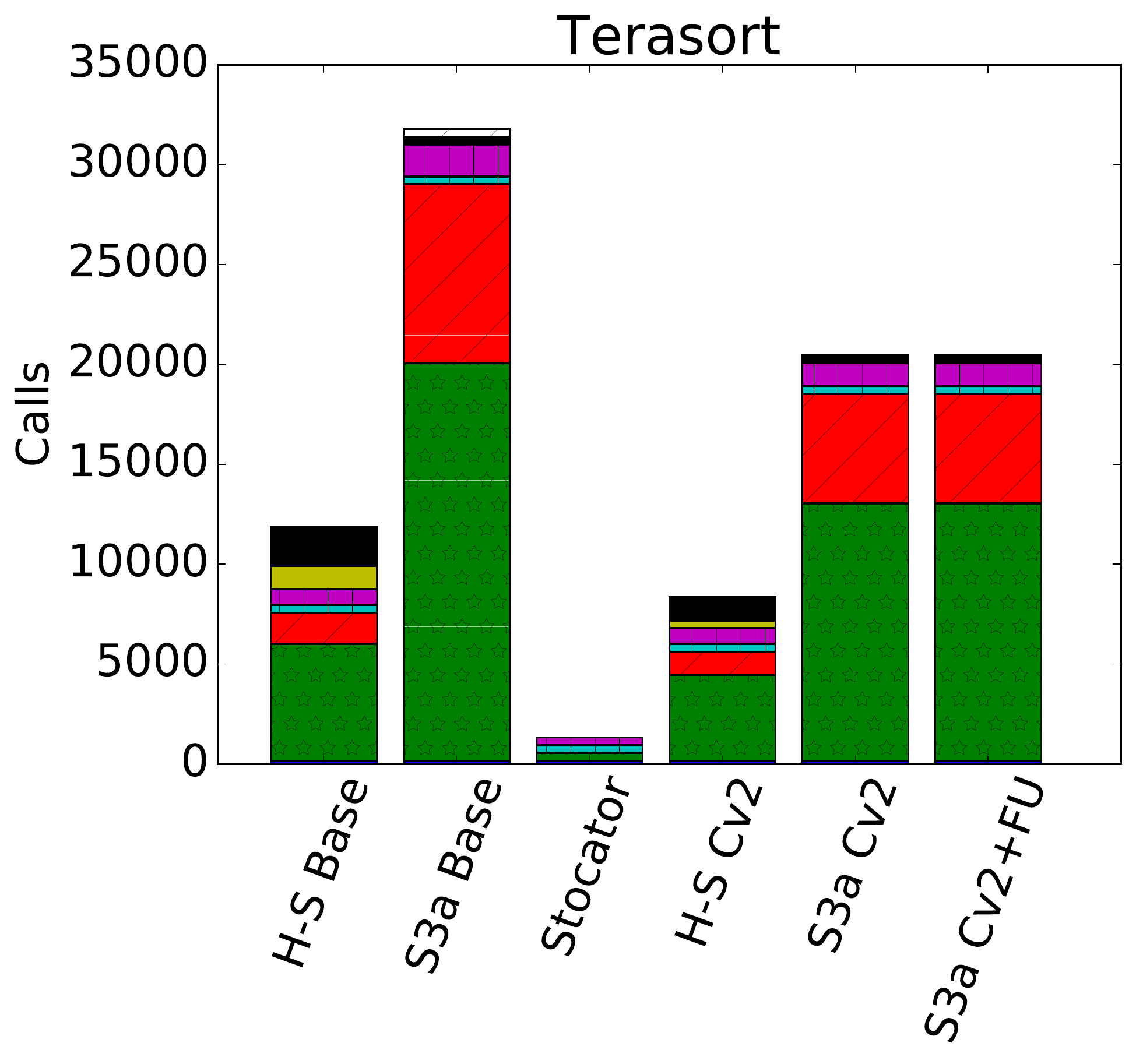}%
	~
	\includegraphics[width=0.3\textwidth]{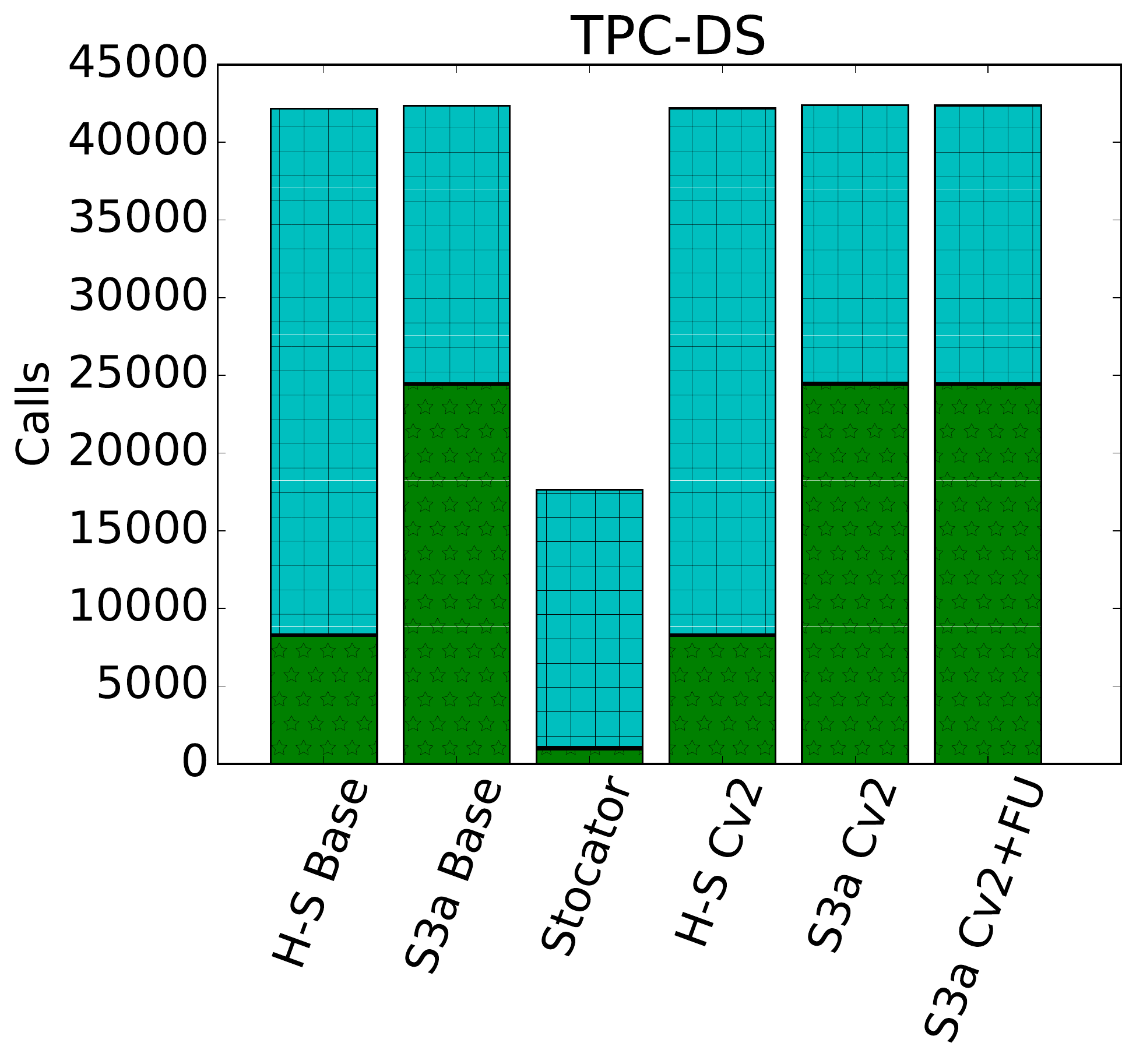}%
	\newline
	\includegraphics[width=0.6\textwidth]{rest/legend.PNG}%
	
	\caption{Macro-benchmarks REST calls comparison}
	\label{fig:macro-rest-comparison}
\end{figure*}

\begin{table*}[t]
	\centering
	\renewcommand{\arraystretch}{1,3}
	\resizebox{0.95\textwidth}{!}{%
		\begin{tabular}{|c|c|c|c|c|c|c|c|}
			\hline 
								& Read-Only 50GB & Read-Only 500GB & Teragen & Copy & Wordcount & Terasort & TPC-DS \\ 
			\hline 
			Hadoop-Swift Base 	& x$2.41$ & x$2.92$  	& x$11.51$ 	& x$9.18$ 	& x$9.21$ 	& x$8.94$ 	& x$2.39$ \\ 
			\hline
			S3a Base 			& x$1.71$ & x$1.96$ 	& x$33.74$ 	& x$24.93$ 	& x$25.35$ 	& x$24.23$ 	& x$2.40$ \\ 
			\hline
			Stocator 			& x$1$ 	 & x$1$ 		& x$1$ 		& x$1$ 		& x$1$ 		& x$1$ 		& x$1$ \\ 
			\hline
			Hadoop-Swift Cv2 	& x$2.41$ & x$2.92$ 	& x$7.72$ 	& x$6.55$ 	& x$6.92$ 	& x$6.29$ 	& x$2.39$ \\ 
			\hline  
			S3a Cv2 			& x$1.71$ & x$1.96$ 	& x$21.15$ 	& x$16.18$ 	& x$16.44$ 	& x$15.41$ 	& x$2.40$ \\ 
			\hline 
			S3a Cv2 + FU 		& x$1.71$ & x$1.96$ 	& x$21.15$ 	& x$16.18$ 	& x$16.44$ 	& x$15.41$ 	& x$2.40$ \\ 
			\hline  
		\end{tabular}
	} 
	\caption{Ratio of REST calls compared to Stocator}
	\label{tab:extra-rest-calls}
\end{table*}

\begin{table*}[t]
	\centering
	\renewcommand{\arraystretch}{1,3}
	\resizebox{\textwidth}{!}{%
		\begin{tabular}{|c|c|c|c|c|c|c|c|}
			\hline 
								& Read-Only 50GB & Read-Only 500GB & Teragen & Copy & Wordcount & Terasort & TPC-DS \\ 
			\hline 
			Hadoop-Swift Base 	& x$9.72$ 	& x$13.67$  & x$8.23$ 	& x$8.60$ 	& x$8.58$ 	& x$8.57$ 	& $2.23$ \\ 
			\hline
			S3a Base 			& x$1.63$ 	& x$1.94$ 	& x$27.82$ 	& x$26.74$ 	& x$26.84$ 	& x$25.88$ 	& $2.25$ \\ 
			\hline
			Stocator 			& x$1$		& x$1$ 		& x$1$ 		& $1$ 		& x$1$ 		& x$1$ 		& x$1$ \\ 
			\hline
			Hadoop-Swift Cv2 	& x$9.72$ 	& x$13.67$ 	& x$5.24$ 	& x$5.86$ 	& x$5.85$ 	& x$5.81$ 	& x$2.23$ \\ 
			\hline  
			S3a Cv2 			& x$1.63$ 	& x$1.94$ 	& x$17.59$ 	& x$17.29$	& x$17.36$ 	& x$16.40$ 	& $2.25$ \\ 
			\hline 
			S3a Cv2 + FU 		& x$1.63$ 	& x$1.94$ 	& x$17.55$ 	& x$17.29$ 	& x$17.34$ 	& x$16.40$ 	& $2.25$ \\ 
			\hline  
		\end{tabular}
	} 
	\caption{REST calls cost compared to Stocator for IBM, AWS, Google and Azure infrastructure}
	\label{tab:relative-costs}
\end{table*}

Next we look at the number of REST operations executed by Spark
in order to understand the load generated on the object storage
infrastructure.
\Cref{fig:micro-rest-comparison,fig:macro-rest-comparison} show that,
in all the workloads, the scenario that uses Stocator achieves the
lowest number of REST calls and thus the lowest load on the object
storage.

When looking at Read-only with
both 50 and 500 GB dataset, the scenario with Hadoop-Swift has the
highest number of REST calls and more than double compared to the
scenario with Stocator. The Hadoop-Swift connector does many more
GET calls on containers to list their contents. Compared to S3a, Stocator
is optimized to reduce the number of HEAD calls on the objects.
We see this consistently for all of the workloads.

In write-intensive workloads, Teragen and Copy, we
see that the scenarios that use S3a as the connector have the highest
number of REST calls while Stocator still has the lowest. Compared to
Hadoop-Swift and Stocator, S3a performs many more HEAD calls for the
objects and GET for the containers. Stocator also does not need to
create temporary \enquote{directories} objects, thus uses far fewer
HEAD requests, and does not need to DELETE objects; this is possible
because our algorithm is conceived to avoid renaming objects after a
task or job completes. \Cref{tab:extra-rest-calls} shows the number of
REST calls that is possible to save by using Stocator.  We observe
that, for write-intensive workloads, Stocator issues 6 to 11 times less REST
calls compared to Hadoop-Swift and 15 to 33 times less compared to S3a,
depending on the optimization features active. 

Having a low load on the Object Storage has advantages both for the
data scientist and the storage providers. On the one hand, cloud
providers will be able to serve a bigger pool of consumers and give
them a better experience. On the other hand, since most public
providers charge fees based on the number of operations performed on
the storage tier, reducing the operations results in a lower cost for
the data scientists.
\Cref{tab:relative-costs} shows the relative costs for the REST operations.
For the workloads with write (Teragen, Copy, Terasort and Wordcount)
Stocator is 16 to 18 times less expensive than S3a run with
FileOutputCommitter version 2, and 5 to 6 times less expensive than
Hadoop-Swift. 
To calculate the cost ratio we used the pricing models of IBM
\cite{ibm-rest-costs}, AWS \cite{aws-rest-costs}, Google
\cite{google-rest-costs} and Azure \cite{azure-rest-costs}; given that
the models are very similar we report the average price.

\begin{figure}[t!]
	\centering
	\includegraphics[width=0.49\columnwidth]{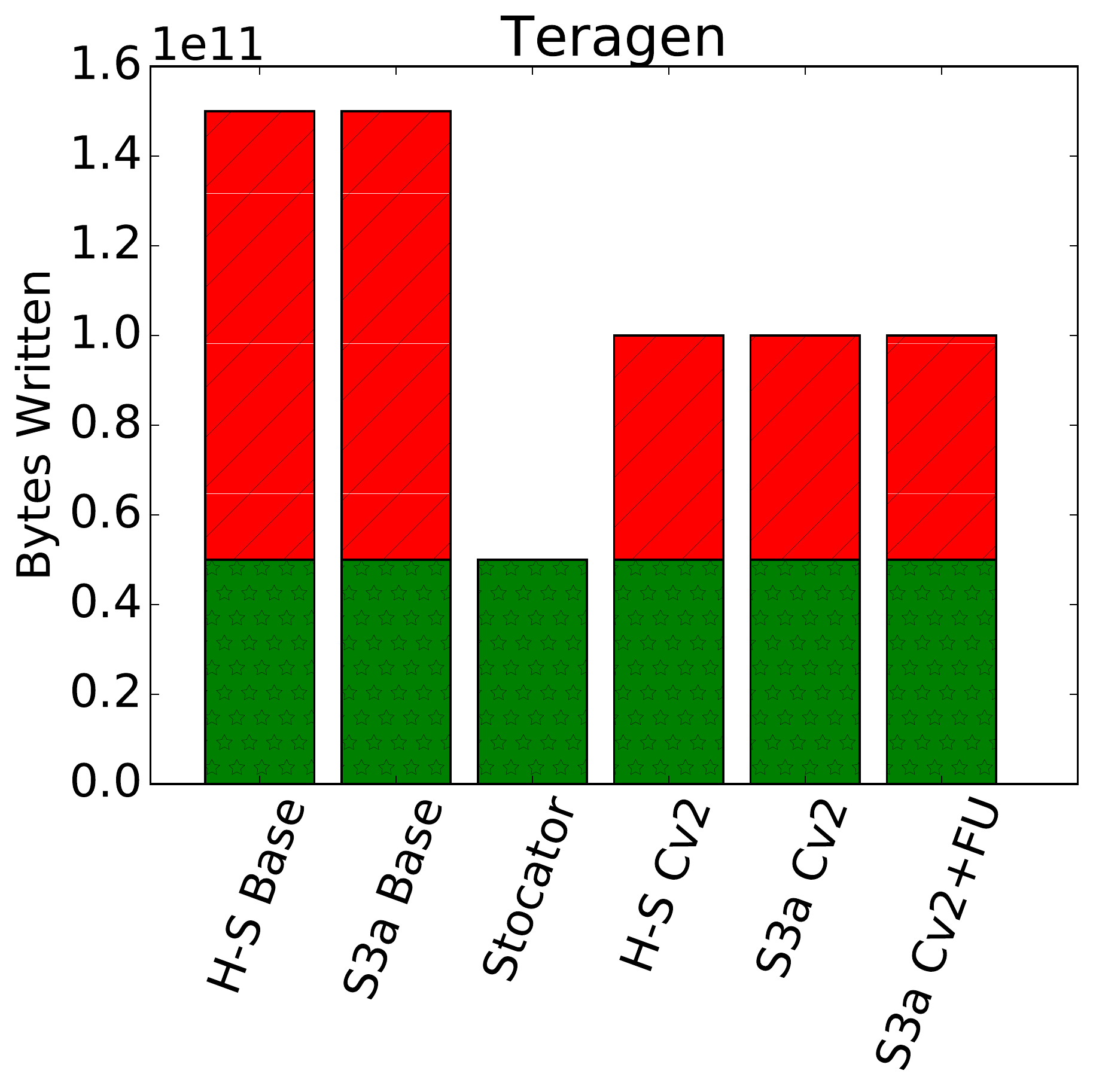}%
	~
	\includegraphics[width=0.49\columnwidth]{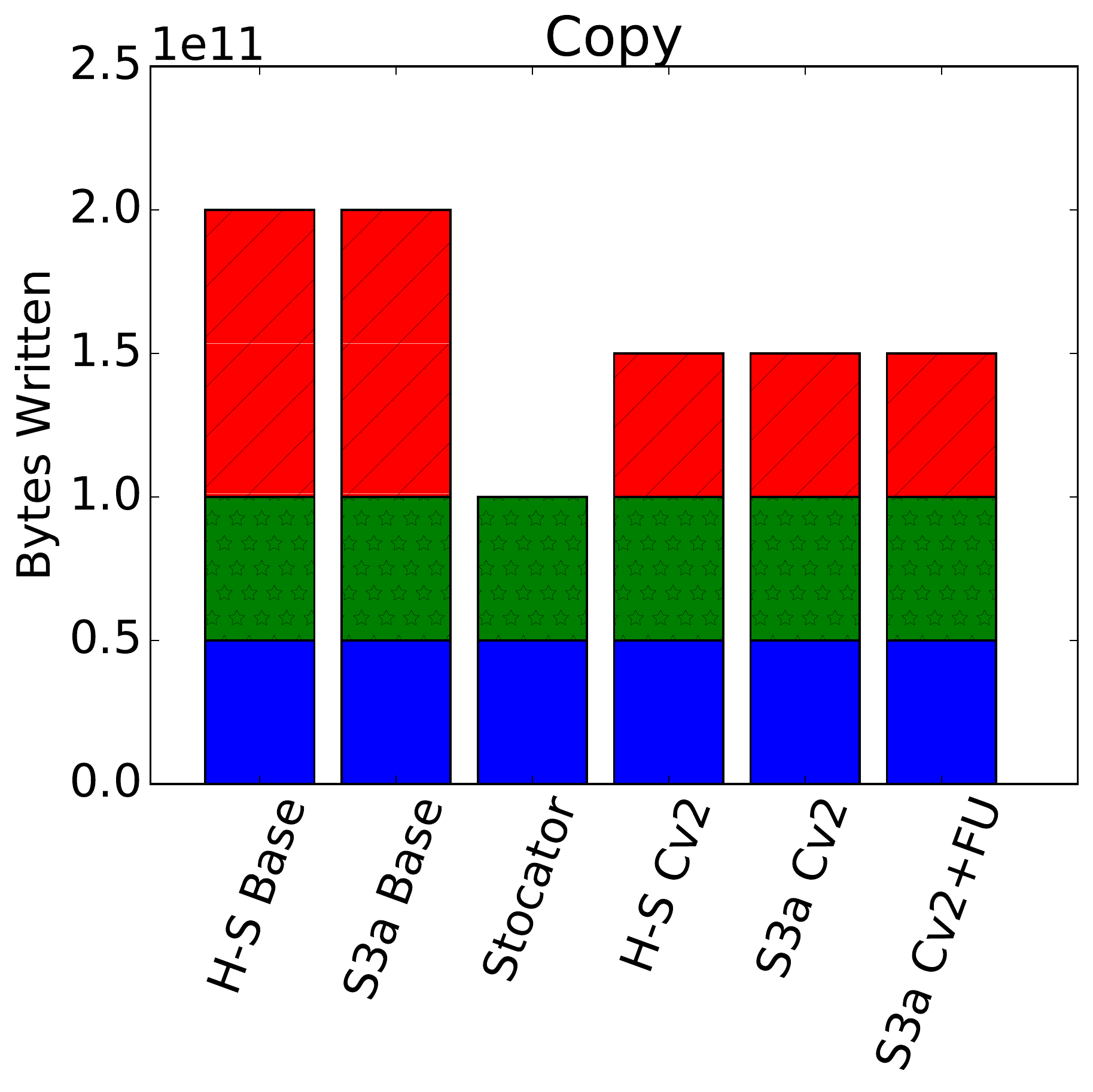}%
	\newline
	\includegraphics[width=0.49\columnwidth]{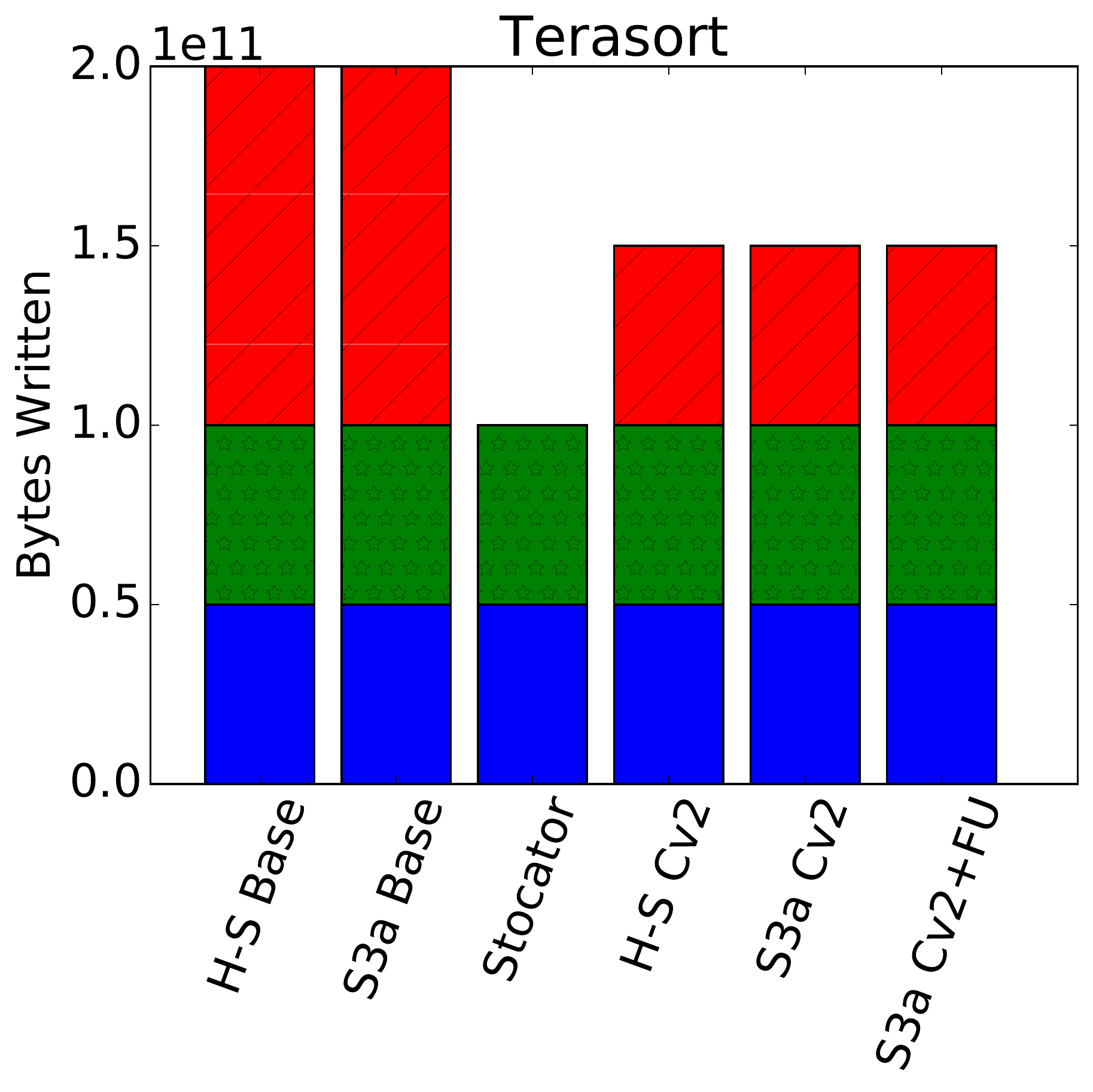}%
	~
	\includegraphics[width=0.49\columnwidth]{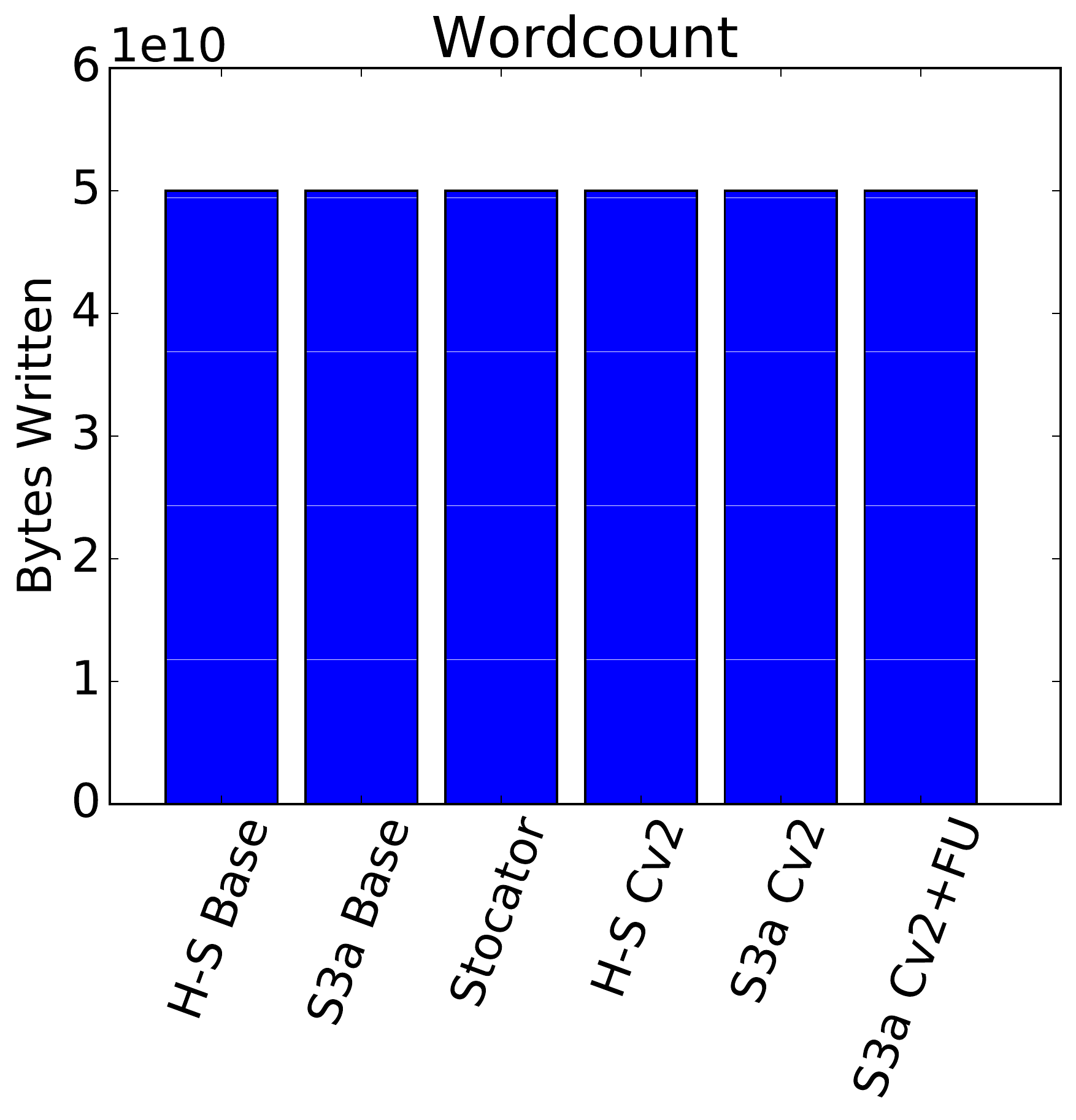}%
	\newline
	\includegraphics[width=0.8\columnwidth]{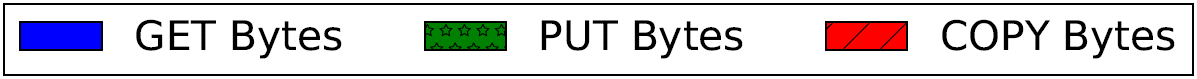}%

	\caption{Object Storage bytes read/written comparison}
	\label{fig:bytes-comparison}
\end{figure}

As an additional way of measuring the load on the object storage and
confirming the fact that Stocator does not perform COPY (or DELETE)
operations we present the number of bytes read and
written to the object storage.
From \cref{fig:bytes-comparison} we see that Stocator does not write
more data than needed on the storage.
In contrast we confirm that Hadoop-Swift and S3a base write each
object three times -- one from the PUT and two from the COPY --
while Stocator only does it once.
Only by enabling FileOutputCommitter Version 2 in Hadoop, it is
possible to reduce the COPY operations to one,
but this is still one more object copy compared to Stocator.
We show only the workloads that have write operations since
during a read-only workload, the number of bytes read from the object
storage are identical for all of the connectors and scenarios (as we
see from the Wordcount workload in \cref{fig:bytes-comparison} where
the number of bytes written is very small).
As expected the S3a scenario that uses
the S3AFastOutputStream optimization gains no benefit
with respect to the number of bytes written to the object storage.

\section{Related Work}
\label{sec:related}

In the past several years there has been a variety of work both from
academia and industry~\cite{kay_2015,anant_2011,night_2012,xie_2010,guo_2012,rang_2002,wang_2009,roman_2015,venzano_2013,rupprecht_2014,trivedi2016ir,pace_2016,vogels_2009,s3mper-blog,s3mper-git,emrfs-blog,emrfs-git,s3guard-issue,s3guard-slides}
that target the performance of analytics frameworks with different
configurations of Compute and Storage layers. We can divide this
work into two major categories that tackle performance analysis and
eventual consistency. 

\textbf{Performance Analysis.} Work
from~\cite{kay_2015,anant_2011,night_2012,xie_2010,guo_2012,rang_2002,wang_2009,roman_2015,venzano_2013,rupprecht_2014}
analyze the performance of analytics
frameworks with different configuration of the Compute and Storage
layer. All this work, albeit valid, base their conclusion on limited
information, workloads and configurations that may not highlight some
problems that exist when analytic applications connect to a specific
Data or Storage layer solution.
In particular Ousterhout et al.~\cite{kay_2015} use an ideal
configuration (Compute and Data layer on the same Virtual Machine),
with limited knowledge of the underlying storage system. With the help
of an analysis performed on network, disk block time and percentages
of resource utilization, such work states that the runtime of analytics
applications is generally CPU-bound rather than I/O intensive. A
recent work~\cite{trivedi2016ir} shows that this is not always true; moving
from a 1Gbps to a 10Gbps network can have a huge impact on the application
runtime. Another work~\cite{pace_2016} shows that is possible to further
improve the run times by eliminating impedance mismatch between
the layers, which can highly affect the run times of such applications; one
in particular when using an Object Storage solution (e.g.; Openstack
Swift~\cite{openstack-swift,swift_2014}) as  the Storage layer. 
Concurrently there has also been some work from industry and open
source to
improve this impedance mismatch. Databricks introduced something
called the DirectOutputCommitter~\cite{direct-output-committer} for S3,
but it failed to preserve the fault tolerance and speculation
properties of the temporary file / rename paradigm. At the same time
Hadoop developed version 2 of the
FileOutputCommitter~\cite{file-output-committer-2}, which renames
files when tasks complete instead of waiting for the completion
(commit) of the entire job. However, this solution does not solve
the entire problem.

\textbf{Eventual Consistency.} Vogels~\cite{vogels_2009} addresses the
relationship between high-availability, replication and eventual
consistency. Eventual consistency guarantees that if no new updates
are made to a given data item,
then
eventually all accesses to that item will return the same value. In
particular, when there is eventual consistency on the list operations
over containers/buckets, current connectors from the Hadoop community
for Swift API~\cite{hadoop-swift} and the S3 API~\cite{aws-s3},
can also lead to failures and incorrect executions.
EMRFS~\cite{emrfs-blog,emrfs-git} from Amazon
and S3mper~\cite{s3mper-blog,s3mper-git} from Netflix overcome
eventual consistency by storing file metadata in
DynamoDB~\cite{aws-dynamodb}, an additional storage system separate from the
object store that is strongly consistent. A similar feature called
S3Guard~\cite{s3guard-issue,s3guard-slides} that also requires an
additional strongly consistent storage system is being developed by
the Hadoop open source community for the S3a connector. Solutions such
as these that require multiple storage systems are complex and can
introduce issues of consistency between the stores. They also add cost
since users must pay for the additional strongly consistent
storage. Our solution does not require any extra storage system.

\section{Conclusion and Future Work}
\label{sec:conclusion}

We have presented a high performance object storage connector
for Apache Spark called Stocator,
which has been made available to the open source community
\cite{stocator-git}.
Stocator overcomes the impedance mismatch of previous open source
connectors with their storage,
by leveraging object storage semantics rather than trying to treat
object storage as a file system.
In particular Stocator eliminates the rename paradigm without
sacrificing fault tolerance or speculative execution.
It also deals correctly with the eventually consistent semantics of
object stores without the need to use an additional consistent storage
system.
Finally, Stocator leverages HTTP chunked
transfer encoding to stream data as it is produced to object storage,
thereby avoiding the need to first write output to local storage.

We have compared Stocator's performance with 
the Hadoop Swift and S3a connectors over a range of
workloads and found that it executes far less operations on
object storage, in some cases as little as one thirtieth.
This reduces the load both for client software and the object
storage service,
as well as reducing costs for the client.
Stocator also substantially increases the performance
of Spark workloads, especially write
intensive workloads, where it is as much as 18 times faster than alternatives.

In the future we plan to continue improving the read performance of
Stocator and extending it to support additional elements of the
Hadoop ecosystem such as MapReduce (which should primarily
require testing) and Hive.


\newpage
\balance

\bibliographystyle{abbrv}
\bibliography{stocator-paper}


%

\end{document}